# Observation of Ferromagnetic Phase in the Second Moiré Band of Twisted MoTe$_2$

Liheng An[1*], Haiyang Pan[1*], Wen-Xuan Qiu[2*], Naizhou Wang[1], Shihao Ru[1,7], Qinghai Tan[3], Xuran Dai[1], Xiangbin Cai[1], Qiuyu Shang[1], Xiufang Lu[1], Hao Jiang[1], Xiaodan Lyu[1], Kenji Watanabe[4], Takashi Taniguchi[5], Fengcheng Wu[2†], Wei-bo Gao[1,6,7†]

[1] *Division of Physics and Applied Physics, School of Physical and Mathematical Sciences, Nanyang Technological University, Singapore 637371, Singapore*

[2]*School of Physics and Technology, Wuhan University, Wuhan 430072, China*

[3] *School of microelectronics, University of Science and Technology of China*

[4] *Research Center for Functional Materials, National Institute for Materials Science, 1-1 Namiki, Tsukuba 305-0044, Japan*

[5] *International Center for Materials Nanoarchitectonics, National Institute for Materials Science, 1-1 Namiki, Tsukuba 305-0044, Japan*

[6] *The Photonics Institute and Centre for Disruptive Photonic Technologies, Nanyang Technological University, Singapore 637371, Singapore*

[7] *Centre for Quantum Technologies, National University of Singapore, Singapore.*

†*Corresponding author. Email:* wufcheng@whu.edu.cn, wbgao@ntu.edu.sg,
*\*These authors contribute equally to this work*

**Abstract**: Flat bands and electron correlation in moiré lattices give rise to many exotic phases, including Mott insulators, superconductivity, and topological states. Within the first moiré band, integer and fractional quantum anomalous Hall effects have been observed in twisted bilayer MoTe$_2$ (tMoTe$_2$) at one hole doping and fractional doping per moiré unit cell, respectively. When the second moiré band is fully hole doped, quantum spin Hall insulator has also been reported in tMoTe$_2$ at a certain twist angle. Exotic topological states together with ferromagnetic (FM) states in the high moiré band can potentially exist as well. In this study, we report the observation of a FM phase in the second moiré band in tMoTe$_2$. The FM phase can be tuned by both the doping level and displacement field. At filling around 2.58 holes per moiré unit cell, the FM phase reaches a Curie temperature of 3.5 K. A large displacement field can suppress the FM phase, like the FM phase at the filling of -1. Our results demonstrate the realization of time-reversal symmetry-breaking states in the higher moiré bands in tMoTe$_2$.

## Main Text

## Background

Stacking layered materials with small twist angles provides a solid-state platform for exploring the interplay of correlation and topology[1-3]. Extremely flat bands quench electron kinetic energy and enhance electron many-body correlation effects, which can be controlled by the twisted angle[4], stacking order[5,6], and displacement field. Robust Chern insulators have been observed in many moiré systems, like twisted bilayer graphene/hBN superlattice[7,8], AB stacked $MoTe_2/WSe_2$[5], $tMoTe_2$ [9-14] and rhombohedral graphene multilayers aligned to hBN[15-17].

Among these, $tMoTe_2$, which hosts an effective honeycomb lattice for low-energy carriers, has emerged as an important platform to study the interplay between band topology and electron correlation. The system at integer filling factor ν = -1 (one hole per moiré unite cell) and fractional fillings -2/3, and -3/5 have been confirmed to exhibit, respectively, integer and fractional quantum anomalous Hall effects both through optical and electrical transport method [10,18]. Many theoretical works have been devoted to exploring the Chern insulating and fractional Chern insulating states[19-27]. Till now, most of the experiment work on $tMoTe_2$ mainly focus on the Chern insulator at filling -1[9] and fractional fillings below -1, like -2/3, and -3/5[10,11,14]. These fractional quantum anomalous Hall states can be understood from composite Fermi Liquid picture based on the mapping from the first moiré Chern band to the lowest Landau level[19,20]. Theories also predict the possibility of the second moiré band mimicking the behavior of the first Landau level[28-30]. Possible ν = −3/2 non-Abelian states are proposed in the second moiré band[29]. A recent experiment has shown the evidence of the existence of quantum spin Hall states[31,32] in fillings -2 and -4. Evidence of Chern insulating states is also reported in twist bilayer $WSe_2$ by finely controlling the twist angle, at fillings -1 and -3[33].

In this study, we report the observation of FM states in the second moiré band, from ν = -2.4 to -2.58. The twist angle of $tMoTe_2$ under study is at 3.2° with moiré density around $n_M \approx 2.7 \times 10^{12}$ $cm^{-2}$. We performed reflective magnetic circular dichroism (RMCD) to investigate the magnetic properties. The RMCD hysteresis loops revealed the existence of ferromagnetism at fillings around both -1 and -2.5. This finding indicates the time reversal symmetry breaking states in higher moiré bands. Furthermore, we investigated the doping dependent and electric field dependent magnetism in the second moiré band.

## Device structure and characterization

The $tMoTe_2$ device is fabricated into dual gate structures as shown in Fig. 1a. We adopt similar tear and stack method[9-11] to fabricate $tMoTe_2$ as used in previous work[9,10]. Figure 1b shows the schematic image of the AA-stacked $tMoTe_2$. In each moiré unit cell, there are three high-symmetry stacking sites as labeled by MM (transition-metal atoms on transition-metal atoms), MX (transition-metal atoms on top of chalcogen atoms), XM (chalcogen atoms on top of transition-metal atoms). For the full filling doping (ν = -2), the holes equally populate on MX and XM sublattices, which form a honeycomb lattice in real space.

Few-layer graphite is used to form both the top gate and bottom gate with hexagonal boron nitride (hBN) as dielectric material, shown in Supplementary Fig. 3d. In addition, the whole twisted bilayer is grounded with another graphite electrode. The voltage sources control the top ($V_{TG}$) and bottom ($V_{BG}$) gate voltages, which allows independent control of both the carrier density and the displacement field. For our devices, small intrinsic doping and built-in displacement field are observed, possibly owing to the imperfection sample fabrication process, which has been calibrated by the dual gate photoluminescence (PL) map measurement and RMCD measurement.

A continuous-wave 780 nm laser is used to excite the tMoTe$_2$ for PL emission[34]. The measurement is kept at 1.6 K unless otherwise specified. The PL signal for emission wavelength above 1064 nm is collected by Superconducting nanowire single-photon detector (SNSPD). As shown in Supplementary Fig. 4d, the real space PL map is measured for the whole device. We select the homogeneous areas to collect the PL signal. The tMoTe$_2$ system with small twist angles like 3-4° is a direct band gap semiconductor[9]. Figure 1d illustrates the PL signal versus the doping level along the yellow dashed linecut in Fig. 1e. We use trion emission as a sensitive probe to characterize incompressible states in tMoTe$_2$. At particular integer fillings like -1 and -2, the free carriers are reduced due to charge gap formation, suppressing the emission of trion. These dips in PL counts correspond to incompressible states at integer fillings like ±2 and ±1.

To verify whether the filling -1 state realizes a Chern insulator, the PL counts are measured as a function of the doping and an applied out-of-plane magnetic field, as presented in Supplementary Fig. 5. The incompressible state at filling -1 disperses linearly with the magnetic field. The Chern number is extracted to be -1, which is consistent with previous work[10].

In Fig. 1e, we measure the PL emission as a function of $V_{TG}$ and $V_{BG}$ and plot it as a function of doping and displacement field. Through the PL map, we deduce the ratio of the top gate and back gate capacitance. With these calibrations and the measurement of hBN thickness (seen in Supplementary Fig. 6), it allows us to identify the twist angle for this device to be around 3.2 ± 0.1°. We use the carrier density difference between integer filling -1 and -2 as moiré density[11].

**The ferromagnetic phase in the second moiré band**

We performed RMCD measurements[35] to probe the magnetism. Figure 2a plots the RMCD intensity as a function of the doping and displacement field. A small out-of-plane magnetic field B = 30 mT is applied to supress the RMCD fluctuation. The signature of FM state is observed in the vicinity of filling -1, ranging from -0.6 to -1.2 and displacement field from +0.12 V/nm to -0.12 V/nm. The RMCD signal on different sample places is shown in Supplementary Fig. 7, which indicates that our device exhibits a homogeneous RMCD signal over a few micrometers. In addition, another filling range with non-zero RMCD is found between filling -2.4 to -2.58, a direct evidence of FM phase formation in the second moiré band.

At fixed carrier densities, we performed RMCD measurement at a larger magnetic field range from $B$ = -0.4 T to 0.4 T. As shown in Fig. 2b, the RMCD intensity is plotted as a function of filling factor and magnetic field. The displacement field is kept at around zero. The RMCD signal is most pronounced in the first moiré band, from filling $v$ = -0.75 to -1.2. For the state at $v = -2$, no RMCD signal is observed, indicating the absence of ferromagnetism. Therefore, the filling -2 state is likely a quantum spin Hall insulator, of which the band structure calculated with Hartree-Fock (HF) approximation is shown in Supplementary Fig. 1b. The RMCD signal at filling -2.5 saturates above 50mT and reach ~ 0.3%. The RMCD signal in the second moiré band is a few times smaller than the signal in the first moiré band. This can be due to different bandwidth and interaction strength in the first and second bands.

To resolve the hysteresis loop, we perform wavelength-dependent RMCD measurement[9]. Based on the PL spectrum of tMoTe$_2$ as shown in Supplementary Fig. 8, we selected the excitation laser wavelength of 1100 nm to 1125 nm for the RMCD measurement. Although the RMCD signal shows different slope based on the wavelength used (Supplementary Fig. 9), an obvious jump of the RMCD signal is observed from -0.05 T to 0.05 T in all cases, indicating the formation of FM states at filling -2.58. Based on these measurements, the wavelength 1121 nm is selected since it gives the largest RMCD contrast.

Figure 2d, e, f and g show the RMCD hysteresis loop at fillings 0, -1, -2 and -2.5. The orange (blue) colour line is for up (down) sweeps. As doping level is set at $v = 0$, no hysteresis loop is resolved and the RMCD signal is almost zero between magnetic field from -40 mT to 40 mT. The coercive field for $v = -1$ is about 10 mT and the difference in RMCD amplitude for up and down sweeps is about 4%. As the doping is increased to around $v = -2$, no hysteresis loop or RMCD signal is detected. When further doping more carriers into the second moiré band, another pronounced hysteresis loop of the RMCD signal is shown, which is a direct evidence of FM state. The coercive field of filling -2.5 is around 10 mT and the RMCD amplitude is around 0.4%, smaller than that of $v = -1$.

The above feature of RMCD signal is consistent with our numerical results. Based on HF approximation, we perform calculation on out-of-plane FM phase for filling -1 to -3. The energy of this phase relative to the symmetric phase is shown in Supplementary Fig. 2, which indicates that the FM phase is favored around filling -1 and -3. As the filling approaches -2 from either side, the ferromagnetism is continuously suppressed. The ferromagnetism at $v = -2.5$ is revealed by the HF band structure, as shown in Fig. 2c, where the second band with spin (valley) polarization is partially filled by holes and contributes to the observed ferromagnetism. Here $v = -2.5$ is a metallic state and has a weaker ferromagnetism compared to the FM Chern insulator $v = -1$.

These hysteresis loops are sensitive to the excitation laser power we used. We kept the laser power at 20 nW to avoid laser heating effects. As shown in Supplementary Fig. 10, the hysteresis loop will gradually disappear when further increasing the power of resonant laser power to 200 nW. This is possibly due to the heating effect that can break down the FM states.

**Doping and electrical control of the ferromagnetic phase**

The FM states can be tuned effectively by doping and displacement field. RMCD signal near $v = -1$ is scanned as a function of filling factor with zero displacement field, as shown in

Supplementary Fig. 11. The filling -1 state gives the strongest coercive field. Similarly, Figure 3a shows RMCD signal as a function of the filling factor $v$ and the out-of-plane magnetic field. The difference of RMCD for the forward and backward magnetic field scanning is shown in Fig. 3b. The coercive field for $v < -2$ shows a maximum at around filling -2.5. RMCD hysteresis loops are shown in Fig. 3c. The FM phase in second moiré band is reproducible in another device#O2 with a twist angle 2.8° (Supplementary Fig. 12). For device#O2, the FM hysteresis loop is most pronounced at filling -2.78 (Supplementary Fig. 13).

We also measure the RMCD signal as a function of the displacement field (Fig. 4). The hysteresis loop has the trend of suppression by the increasing of displacement field. The coercive field gradually decreases from 1.5 mT at electric field 13 mV/nm to 0.5 mT at 47 mV/nm. An unsymmetric coercive field is observed for all the electric field as shown in Fig. 4c, of which the microscopic origin still needs more investigation. The electric field dependence is also observed in Device#O2 (Supplementary Fig. 14).

**Cuire temperature of the FM phase**

To investigate the magnetic interaction strength, we measure RMCD signal as a function of temperature. At a low temperature T = 1.6 K, a clear RMCD signal jump can be detected, which indicates FM interaction between spin moments at filling -2.58. For the high temperature like 7K, the RMCD curve shows a linear relation with the magnetic field, which is the behaviour of the paramagnetic phase. We extract the slope of RMCD to calibrate the magnetic susceptibility $\chi = \frac{\partial R}{\partial H}|_{H=0}$[9]. We fit the inverse of the slope in the paramagnetic phase as a function of the temperature following the Cuire-Weiss law described by $\chi = \frac{C}{T-\theta_C}$, as shown in Fig. 5b. By fitting the curve, a positive $\theta_C = 3.5$ K is extracted, which is the FM Curie temperature.

The temperature dependence of RMCD signal is also measured with optimal excitation laser wavelength 1120.9 nm for $v$ below -2.4, as shown in Fig. 5c. We measure the RMCD signal by sweeping the magnetic forward and backward from $T$ = 1.6 K to 4.5 K. At this wavelength, the RMCD signal shows hysteresis loop that gradually disappears as the temperature increases. The temperature at which the hysteresis loop vanishes leads to a Curie temperature of $T_c \sim 3.5$ K at $v$ = -2.58, consistent with the above estimate. The Curie temperature obtained in this way shows doping dependence when the filling is tuned from -2.4 to -2.58, which gradually decreases when $v$ deviates from -2.58.

**Outlook and conclusion**

Previous work in small twist angle 2.1° has shed light to realize fractional quantum spin Hall insulating states in second moiré band in tMoTe$_2$[36] at $v = -3$. The investigation of high order moiré band opens up the possibility of realizing more topological states, including non-Abelian states[28,29,32,37,38]. Our study demonstrated that FM phase can also appear in the second moiré band and is tunable by both doping level and the displacement field. As can be expected in the near future, different twist angles and multi-layer structures can lead to other possible topological phases, making the second moiré band as an interesting place to study both many-body correlation and band topology.

**Methods**

Device fabrication

The dual-gate tMoTe$_2$ was fabricated using a method similar to that previously reported [10,11], as shown in Supplementary Fig. 3. Bulk crystal MoTe$_2$ was purchased from HQ Graphene. First, few-layer graphite and hBN (10 to 30 nm), as shown in Supplementary Fig. 6, were mechanically exfoliated onto a silicon wafer. The cleanness of the top surface was checked using both an optical microscope with dark field and Atomic Force Microscopy (AFM). Polydimethylsiloxane (PDMS) and Polycarbonate (PC) were used to pick up the bottom hBN (the bottom substrate) and bottom graphite (the bottom gate). The whole stack was annealed at 200°C in a forming gas atmosphere to eliminate residues at the interface. The exfoliation and transfer process were performed in a nitrogen-filled glovebox, where the water and oxygen concentrations were kept below 0.1 ppm. Monolayer MoTe$_2$ was mechanically exfoliated onto a Si/SiO2 wafer using scotch tape. The monolayer MoTe$_2$ was cut in half with an AFM tip. Then, another layer of top graphite (5-10 layers) /top hBN (10 to 30 nm) stack was used to pick up the two parts of MoTe$_2$ sequentially. The twist angle was set by rotating the bottom transfer stage by 3°. Subsequently, another 5 to 10 layers of graphite were picked up and contacted with the bottom MoTe$_2$ to form the ground. The whole stack was dropped onto the pre-patterned bottom electrode (Cr/Au, 5/30 nm) to form the complete electric contact. The PC film on top was washed with a chloroform rinse followed by an isopropyl alcohol (IPA) rinse.

Optical measurements set-up

All optical measurements were performed in a home-built confocal optical microscope. The sample was mounted in the cryostat Autodry 2100 equipped with a 9T superconducting magnet at base temperature 1.6K. A × 50 non-magnetic objective (numerical aperture, 0.63) with a spot size of 1um was used to collect PL and RMCD signal. For the PL measurement, a M-squared laser with tunning range 700nm-1000nm was used as the excitation source. A 1064nm long pass filter is placed before the entrance to the spectrometer to remove the reflected laser excitation. The PL excitation power ranges from 200nW to 2uW and the gate dependent PL counts signal is obtained by superconducting single-photon detector (SNSPD). Princeton spectrometer equipped with a liquid-nitrogen-cooled InGaAs charge-coupled device (CCD) detector is used to analyse the PL spectrum.

The reflective magnetic circular dichroism (RMCD) measurement was performed with a continuous wave laser (HUBNER Photonics C-WAVE VIS), which produced a narrow MHz bandwidth laser, ranging 1080nm to 1130nm. The set up is shown in Supplementary Fig. 15. The excitation laser was first passed through an optical chopper at frequency $f_1$ 912Hz and then polarized 45° to the photoelastic modulator (PEM) fast axis with a maximum retardance of λ/4. The departing circularly polarized light oscillates between right circular polarized light and left circular polarized light in a sinusoidal time-dependence with a frequency $f_2$ 50.5kHz. The reflected signal from the sample is reflected by a beamsplitter and sent into an InGaAs avalanche photodiode (Thorlabs APD430C/M). The photodiode current was amplified and turned into voltage signal by SR570 and sent into two lock-in amplifiers set at 50.5kHz and 912Hz, giving the RMCD signal and the laser excitation intensity respectively.

Carrier density calculation

The carrier density is $n = (V_{TG}C_{TG} + V_{BG}C_{BG})/e - n_0$ and displacement field $D/\varepsilon_0 = (V_{TG}C_{TG} - V_{BG}C_{BG})/2\varepsilon_0 - D_0/\varepsilon_0$, where $C_{TG}$ and $C_{BG}$ are the top and bottom gate capacitance obtained from the device geometry, $e$ is the electron charge, and $\varepsilon_0$ is the vacuum permittivity. For device#O1, the top BN thickness is 19nm, and $C_{TG}/e = \varepsilon\varepsilon_0/d = 0.87\times10^{12}$ cm$^{-2}$. The bottom BN thickness is 24nm, and $C_{BG}/e = \varepsilon\varepsilon_0/d = 0.69\times10^{12}$ cm$^{-2}$. For device#O2, the top BN thickness is 44nm, and $C_{TG}/e = \varepsilon\varepsilon_0/d = 0.38\times10^{12}$ cm$^{-2}$. The bottom BN thickness is 40nm, and $C_{BG}/e = \varepsilon\varepsilon_0/d = 0.41\times10^{12}$ cm$^{-2}$. The offset carrier density $n_0$ is derived from fitting the integer filling in dual gate PL spectra map. The offset displacement field $D_0$ is determined from the symmetric axis of the dual gate RMCD map.

Note added.—Recently, two complementary and independent study[39,40] appeared on Arxiv.

**Acknowledgement**

We thank Jinxin Hu, Justin Song, Bo Yang, Jiaqi Cai, Xiaodong Xu for helpful discussions. This work is supported by ASTAR (M21K2c0116, M24M8b0004), Singapore National Research foundation (NRF-CRP22-2019-0004) and CQT bridge funds. W.-X. Q. and F.W are supported by National Key Research and Development Program of China (Grants No. 2021YFA1401300 and No. 2022YFA1402401), National Natural Science Foundation of China (Grant No. 12274333). W.-X. Q. is also supported by the China Postdoctoral Science Foundation (Grants No. 2024T170675 and No. 2023M742716). The numerical calculations in this paper have been done on the supercomputing system in the Supercomputing Center of Wuhan University.

**Author contributions**

H.P. fabricated the devices. L.A. performed the optical measurements. L.A., H.P., F.W., W.G. analyzed the data. W.-X. Q. and F.W performed theoretical studies. K.W. and T.T. grew the bulk hBN crystals. L.A. and W.G. together with all coauthors wrote the manuscript. All authors discussed the results and commented on the manuscript.

**Competing interests**

The authors declare no competing financial interests.

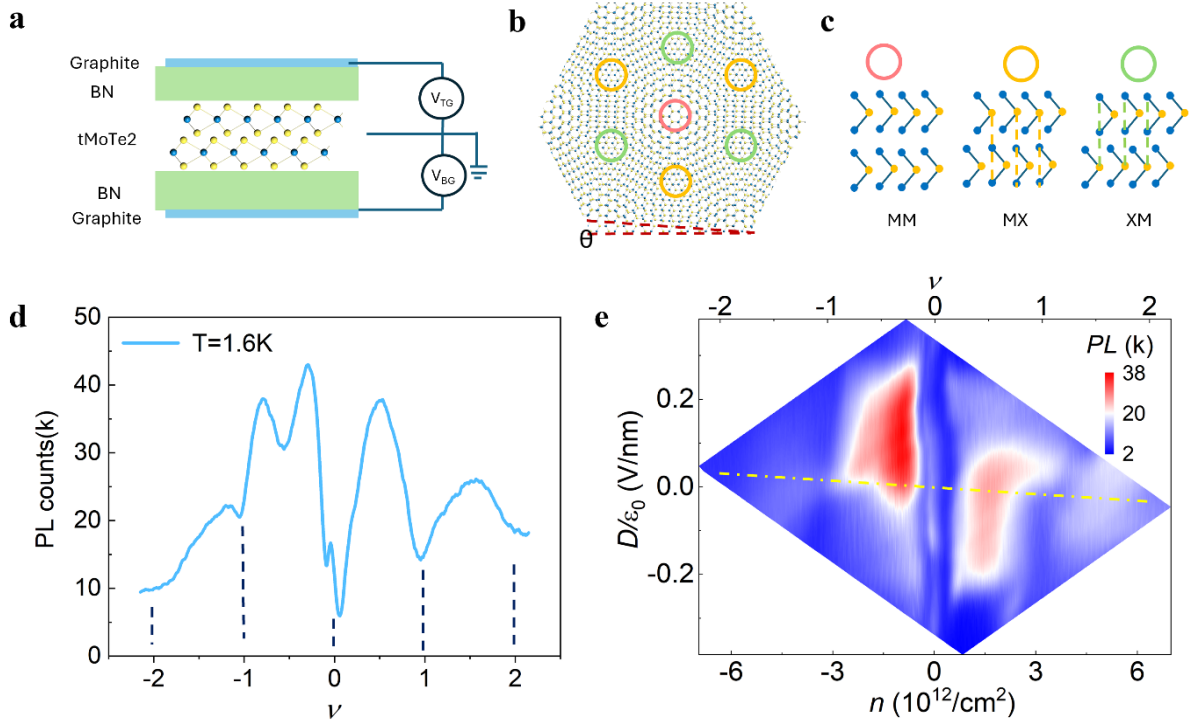

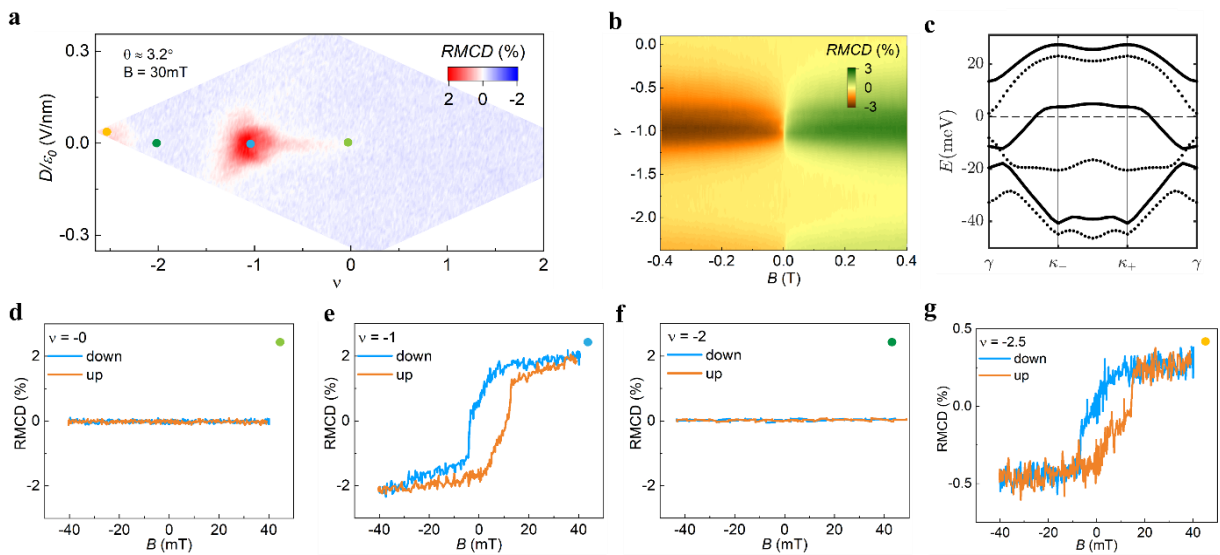

**Fig. 1 | Gate-tunable correlated states in tMoTe2 devices. a.** Schematic image of the device structure. The tMoTe$_2$ sample is grounded. The top gate and bottom gate are applied through hBN (10 to 30 nm). **b.** Moiré superlattice of AA stacked tMoTe$_2$. The red, orange, and green circles mark high-symmetry stacking sites MM, MX, and XM, respectively. **c.** Side view of MM, MX, and XM sites. **d.** PL counts as a function of doping level, with displacement field ($D/\varepsilon_0$) almost at zero, along the linecut of the yellow dashed line shown in Fig. 1e. **e.** Steady state photoluminescence (PL) versus carrier density (filling factor) and displacement field.

**Fig. 2 | Ferromagnetism in the first and second moiré band. a.** Reflective magnetic circular dichroism (RMCD) signal as a function of the doping filling factor $\nu$ and electric field $(D/\varepsilon_0)$.

The wavelength is at 1120.9 nm. **b.** RMCD signal as a function of out-of-plane magnetic field $B$ and filling factor $v$ from 0 to -2.4. **c.** Calculated Hartree-Fock (HF) band structure at $v = -2.5$ in the metallic FM phase. The solid and dotted lines, respectively, plot spin up and down bands. The Fermi energy marked by the horizontal black dashed line is set to 0. **d. e. f.** RMCD signal at filling factor 0, -1 and -2 respectively, with the displacement field near $D/\varepsilon_0 = 0$ mV/nm. **g.** RMCD signal at filling factor -2.5, with the displacement field $D/\varepsilon_0 \approx 10$ mV/nm.

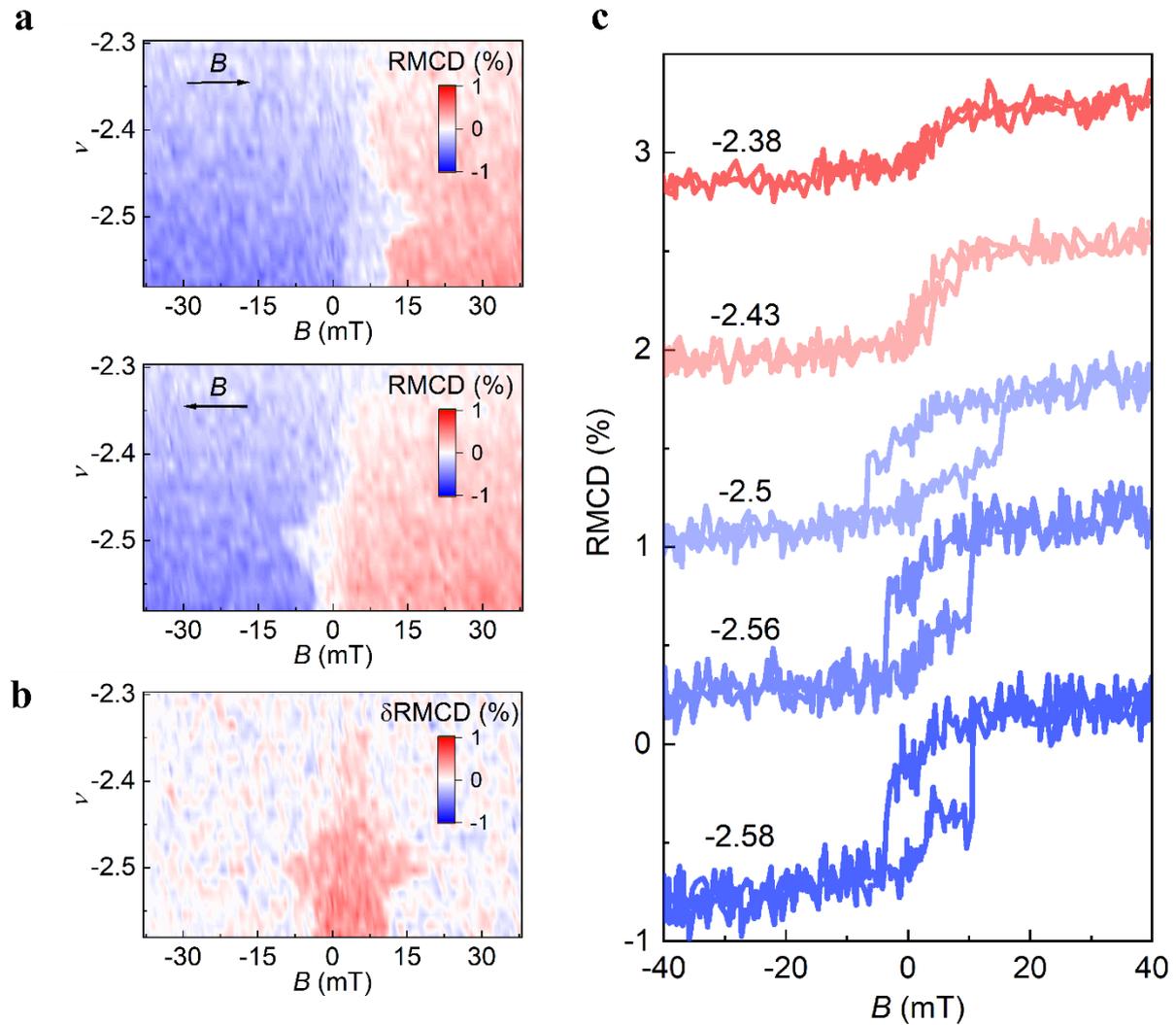

**Fig. 3 | Doping dependent RMCD from filling -2.3 to -2.58. a.** RMCD signals versus filling and magnetic field swept up and swept down. **b.** Hysteresis loop of the RMCD as a function of magnetic field and filling factor. **c.** RMCD signal as a function of magnetic field from filling -2.38 to -2.58.

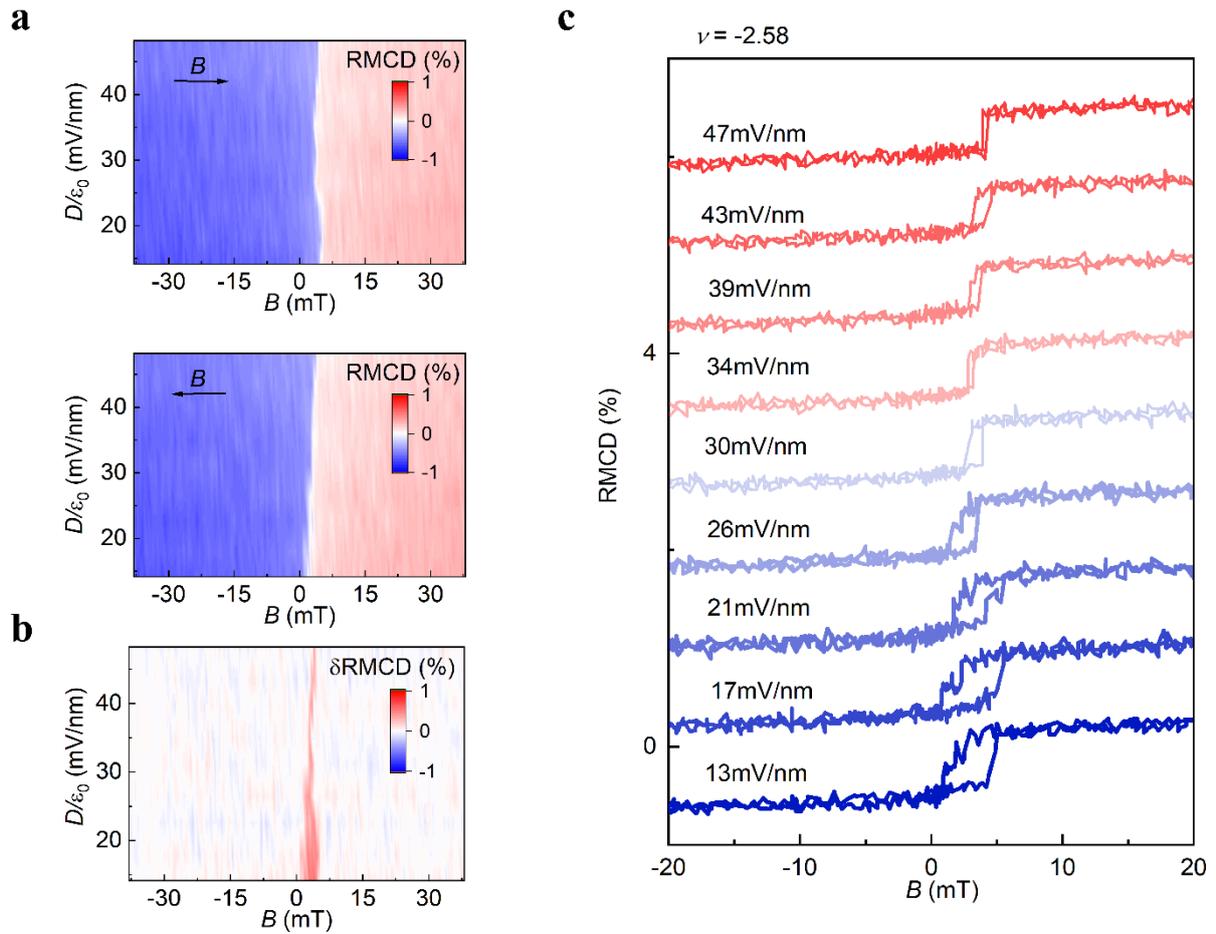

**Fig. 4 | Electric filed dependent RMCD at filling -2.58. a.** RMCD signals versus displacement field and magnetic field swept up and swept down. **b.** Hysteresis loop of the RMCD versus displacement field. **c.** RMCD signal at filling -2.58 with displacement field from 13 mV/nm to 47 mV/nm.

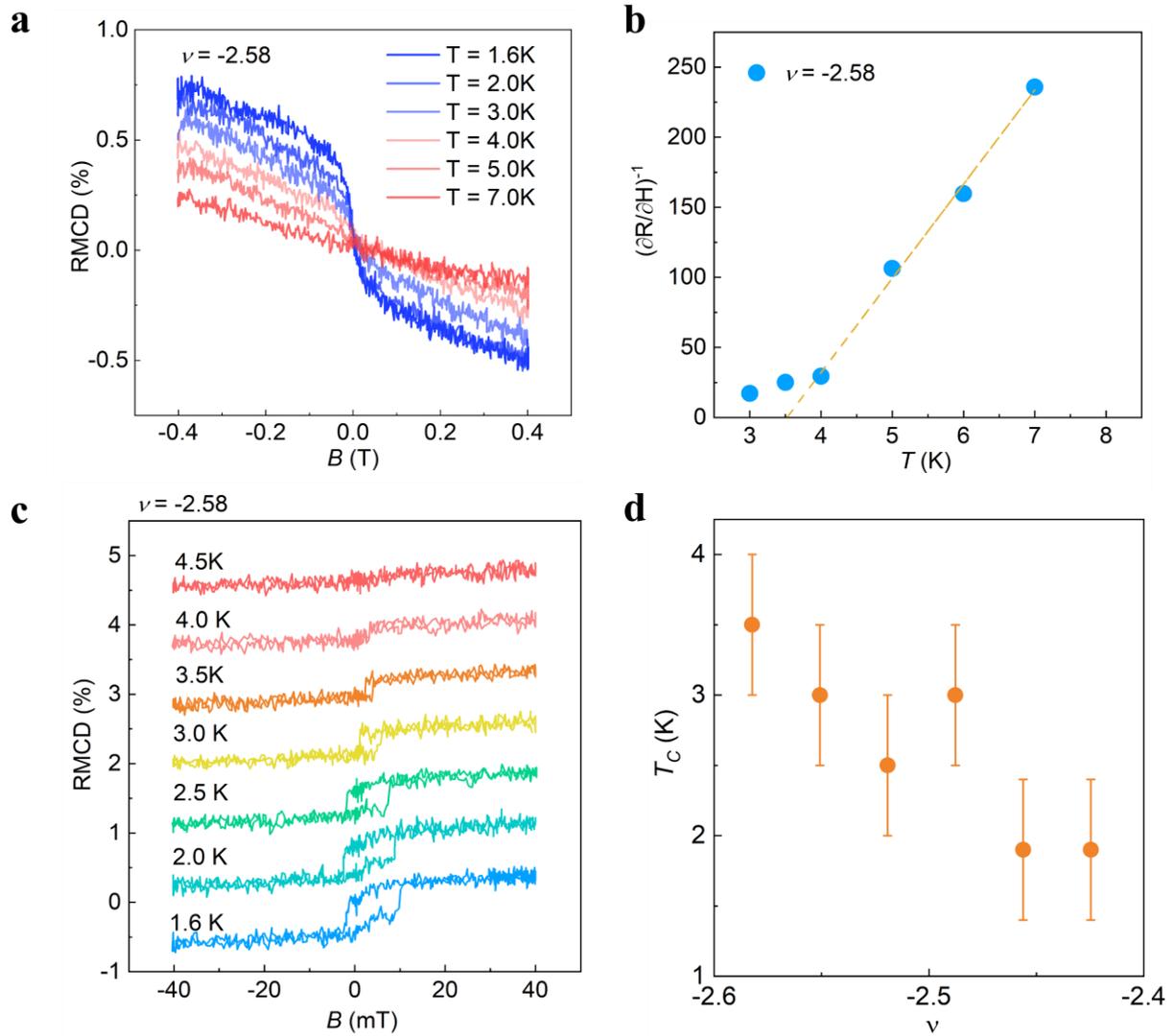

**Fig. 5 | $T_C$ of ferromagnetic phase at filling -2.58. a.** Temperature dependent RMCD at filling factor $v = -2.58$. The measurement wavelength is at 1104.7 nm. **b.** Curie-Weiss fit (dashed line) of the inverse RMCD slope at vanishing magnetic field versus temperature, which leads to $T_C$ of 3.5 K. **c.** Temperature dependent RMCD hysteresis at filling factor $v = -2.58$. The measurement wavelength is at 1120.9 nm. **d.** Curie temperature, at which the RMCD hysteresis loop vanishes, as a function of filling factor.

# Supplementary Information

# Observation of Ferromagnetic Phase in Second Moiré Band of twisted MoTe$_2$

Liheng An[1*], Haiyang Pan[1*], Wen-Xuan Qiu[2*], Naizhou Wang[1], Shihao Ru[1,7], Qinghai Tan[3], Xuran Dai[1], Xiangbin Cai[1], Qiuyu Shang[1], Xiufang Lu[1], Hao Jiang[1], Xiaodan Lyu[1], Kenji Watanabe[4], Takashi Taniguchi[5], Fengcheng Wu[2†], Wei-bo Gao[1,6,7†]

[1] Division of Physics and Applied Physics, School of Physical and Mathematical Sciences, Nanyang Technological University, Singapore 637371, Singapore

[2] School of Physics and Technology, Wuhan University, Wuhan 430072, China

[3] School of microelectronics, University of Science and Technology of China

[4] Research Center for Functional Materials, National Institute for Materials Science, 1-1 Namiki, Tsukuba 305-0044, Japan

[5] International Center for Materials Nanoarchitectonics, National Institute for Materials Science, 1-1 Namiki, Tsukuba 305-0044, Japan

[6] The Photonics Institute and Centre for Disruptive Photonic Technologies, Nanyang Technological University, Singapore 637371, Singapore

[7] Centre for Quantum Technologies, National University of Singapore, Singapore.

†Corresponding author. Email: wufcheng@whu.edu.cn, wbgao@ntu.edu.sg
*These authors contribute equally to this work

**Model Hamiltonian and mean-field calculations.**

I. Moiré Hamiltonian

The single-particle moiré Hamiltonian of tMoTe$_2$ can be constructed for valence band states in ±K valley as[1],

$$\hat{\mathcal{H}}_0^\tau = \begin{pmatrix} \dfrac{-\hbar^2(\hat{k}-\tau\kappa_+)^2}{2m^*} + \Delta_+(r) & \Delta_{T,\tau}(r) \\ \Delta_{T,\tau}^\dagger(r) & \dfrac{-\hbar^2(\hat{k}-\tau\kappa_-)^2}{2m^*} + \Delta_-(r) \end{pmatrix}$$

$$\Delta_\pm(r) = 2V \sum_{j=1,3,5} \cos(g_j \cdot r \pm \psi),$$

$$\Delta_{T,\tau}(r) = w\left(1 + e^{-i\tau g_2 \cdot r} + e^{-i\tau g_3 \cdot r}\right),$$
(1)

where $\hat{\mathcal{H}}_0^\tau$ is expressed in the layer-pseudospin space at valley $\tau$. Here, the index $\tau = \pm K$ are also locked to spin ↑ and ↓. In the moiré Hamiltonian, $\kappa_\pm = [4\pi/(3a_M)](-\sqrt{3}/2, \mp 1/2)$ and $g_j = [4\pi/(\sqrt{3}a_M)]\{\cos[(j-1)\pi/3], \sin[(j-1)\pi/3]\}$. for $j = 1,\ldots,6$, where $a_M \approx a_0/\theta$ is the moiré period, $a_0$ is the monolayer lattice constant, and $\theta$ is the twist angle. Here, we take the parameter values estimated for $t$MoTe$_2$ from Ref [1], $a_0 = 3.472$Å, $m^* = 0.62 m_e$, $V = 8$ meV, $\psi = -89.6°$, $\omega = -8.5$ meV, where $m_e$ is the electron bare mass.

## II. Band-projected Interaction Hamiltonian

We construct the interacting model within hole basis, where the corresponding hole operator is defined as $b_{k n \tau} = c_{k n \tau}^\dagger$, where $c_{k n \tau}^\dagger$ is the creation operator for the Bloch state $\phi_k^{n\tau}(r)$ with energy $\varepsilon_k^{n\tau}$ of the $n$-th moiré band at $k$ and $\tau$. The single-particle Hamiltonian $\hat{\mathcal{H}}_0$ in the hole basis can be written as

$$\hat{\mathcal{H}}_0 = \sum_{k,\tau,n} \mathcal{E}_k^{n\tau} b_{k n \tau}^\dagger b_{k n \tau},$$
(2)

where $\mathcal{E}_k^{n\tau} = -\varepsilon_k^{n\tau}$ is the energy of Bloch state. The projected Coulomb interaction is expressed as

$$\hat{\mathcal{H}}_{\text{int}} = \frac{1}{2} \sum V_{k_1 k_2 k_3 k_4}^{n_1 n_2 n_3 n_4}(\tau,\tau') b_{k_1 n_1 \tau}^\dagger b_{k_2 n_2 \tau'}^\dagger b_{k_3 n_3 \tau'} b_{k_4 n_4 \tau},$$
(3)

where the summation is over the momentum $k_j$ (summed over the moiré Brillouin zone), the moiré band index $n_j$, and the valley index $\tau$. The Coulomb matrix element is given by,

$$V_{k_1 k_2 k_3 k_4}^{n_1 n_2 n_3 n_4}(\tau,\tau') = \frac{1}{\mathcal{A}} \sum_q V_q M_{k_1 k_4}^{n_1 n_4}(\tau, q) M_{k_2 k_3}^{n_2 n_3}(\tau', -q),$$
(4)

$$M_{k_1 k_4}^{n_1 n_4}(\tau, q) = \sum_l \int dr\, e^{iq \cdot r} \left[\tilde{\phi}_{k_1 l}^{n_1 \tau}(r)\right]^* \left[\tilde{\phi}_{k_4 l}^{n_4 \tau}(r)\right],$$
(5)

Where $\mathcal{A}$ is the system area, and $\tilde{\phi}_k^{n\tau}(r) = [\tilde{\phi}_k^{n\tau}(r)]^*$ due to the particle-hole transformation and $l$ is the layer index. We use gate-screened Coulomb potential $V_q = 2\pi e^2 \tanh(|q|d)/(\epsilon|q|)$, where $d$ is the gate-to-sample distance and $\epsilon$ is the dielectric constant. In our calculation, we set $d = 20$ nm, $\epsilon = 20$. The full Hamiltonian is $\hat{\mathcal{H}} = \hat{\mathcal{H}}_0 + \hat{\mathcal{H}}_{\text{int}}$.

## III. Mean-field Calculation

We perform self-consistent calculations based on Hartree-Fock approximation to study possible ground state of $\hat{\mathcal{H}}$ as a function of ν for θ = 3.2°. We keep three moiré bands per valley in the band projection. A 30 × 30 *k*-mesh is used in our calculation and a finite temperature $T = 1.6$ K is assumed. Motivated by the experimental observation, we consider two types of states, (1) a symmetric state without any symmetry breaking; (2) an out-of-plane ferromagnetic state. We self-consistently calculate mean-field solutions for these two types of states. The ground state is then determined by comparing their energies for $\upsilon$ between -1 to -3. As shown in **Supplementary Fig. 1,** ferromagnetism is found from ν =-1 to -1.6 in the first moiré band and from -2.1 to -3 in the second moiré band, which is qualitatively consistent with the experimental findings. We note that there can be other competing states[2]. Here we focus on the symmetric state and the ferromagnetic state to illustrate the main physics. We show representative mean-field band structures in **Supplementary Fig. 2** for (1) ferromagnetic Chern insulator at $\upsilon = -1$, (2) the quantum spin Hall insulator at $\upsilon = -2$, and (3) the ferromagnetic metal at $\upsilon = -2.5$.

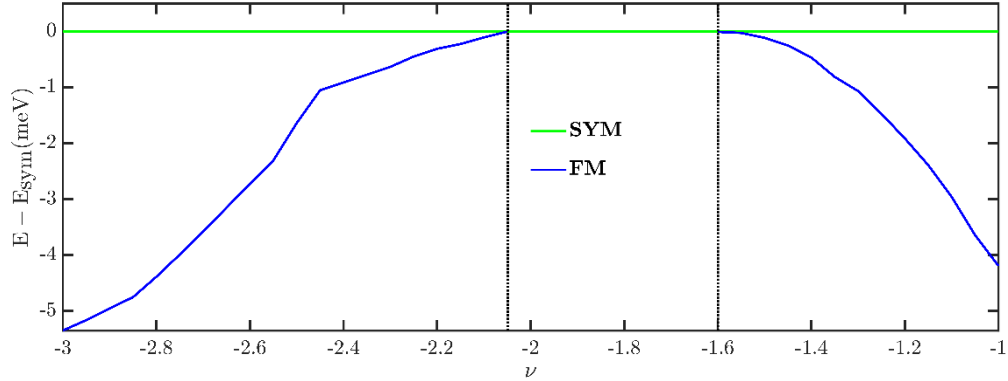

**Supplementary Fig. 1. Energy of out-of-plane ferromagnetic states and symmetric states as a function of filling $\upsilon$ between -1 to -3**. $\theta = 3.2°$. Vertical Dotted lines mark the phase transitions.

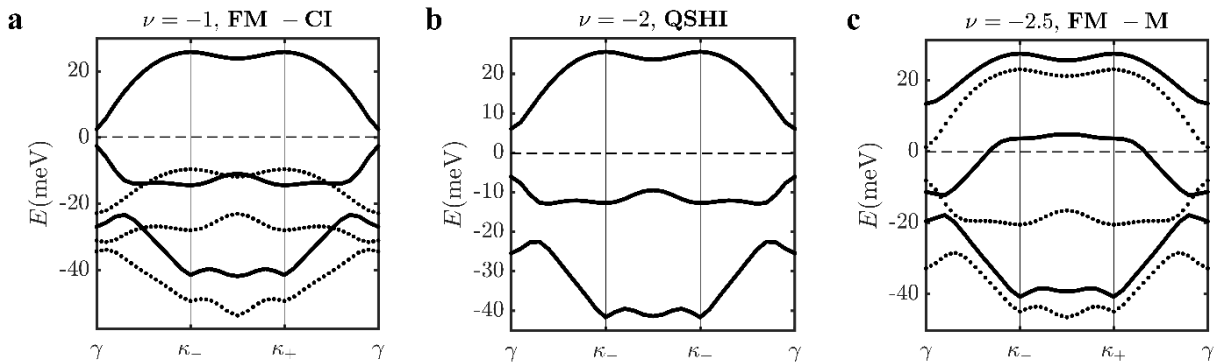

**Supplementary Fig. 2. The mean-field band structure at twist angle $\theta = 3.2°$. a**. The HF band structure at $\upsilon = -1$ is in the ferromagnetic Chern insulator (FM-CI) phase. The middle

of the interaction-induced gap, marked by the black dashed line, is set to 0. The solid and dotted lines, respectively, plot bands from +K valley (spin up) and −K valley (spin down). **b**. The HF band structure at $\nu = -2$ is in the quantum spin Hall insulator (QSHI) phase. The middle of the interaction-induced gap, marked by the black dashed line, is set to 0. The bands are spin degenerate. **c**. The HF band structure at $\nu = -2.5$ is in the ferromagnetic metallic (FM-M) phase. The solid and dotted lines, respectively, plot bands from +K and −K valleys. The Fermi energy marked by the black dashed line, is set to 0. All the band structures are presented in the electron basis defined by $c^\dagger_{kn\tau}$ and $c_{kn\tau}$ operators.

**Device fabrication**

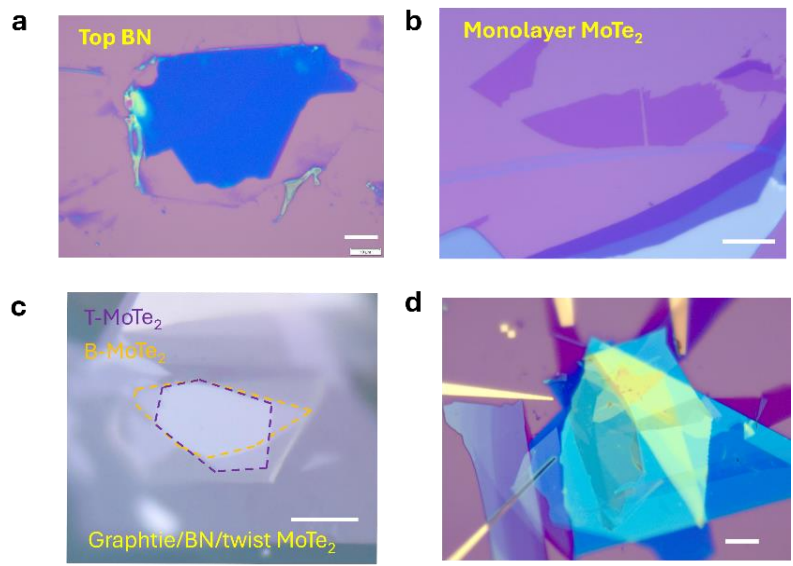

**Supplementary Fig. 3. The fabrication process of our device. a.** The top hBN and monolayer MoTe$_2$ were mechanically exfoliated onto the 300 nm SiO$_2$ with scotch tape. **b.** The monolayer MoTe$_2$ is cut by the AFM tip in glove box. **c.** Image taken after stacking the twist MoTe$_2$ together. The purple colour marks the top layer MoTe$_2$ and the orange colour marks the bottom MoTe$_2$. **d.** Device#O1 on 300 nm SiO$_2$, the electrical connection is through the outside prepatterned electrode Cr/Au, with a thickness 5 nm/30 nm. The scale bar is 10 µm.

In our devices, the top gate, bottom gate, and ground lead are connected to the prepatterned gold leads outside via the few-layer graphite films.

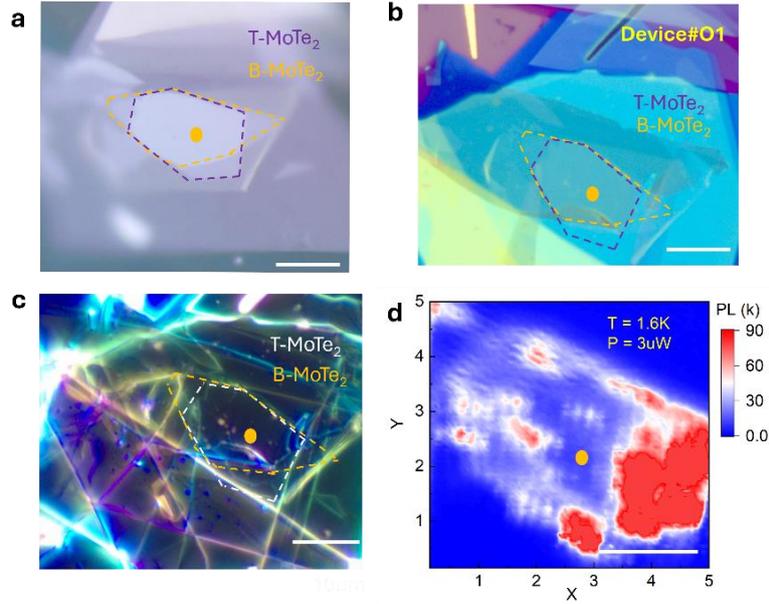

**Supplementary Fig. 4. Optical characterization of Device#O1. a.** The optical image of the twisted bilayer MoTe$_2$ on Polycarbonate (PC). **b**. The optical image of the twist bilayer MoTe$_2$ on SiO$_2$. **c**. Dark field image of the twisted bilayers MoTe$_2$. **d**. The PL counts collected by Superconducting nanowire single-photon detector, with a 1064 nm long pass filter, excitation with 780 nm laser power P = 3μW, T = 1.6 K. Orange point is the area to collect the PL and RMCD signal. The scale bar is 10 μm.

**The PL map as a function of doping and out-of-plane magnetic field**

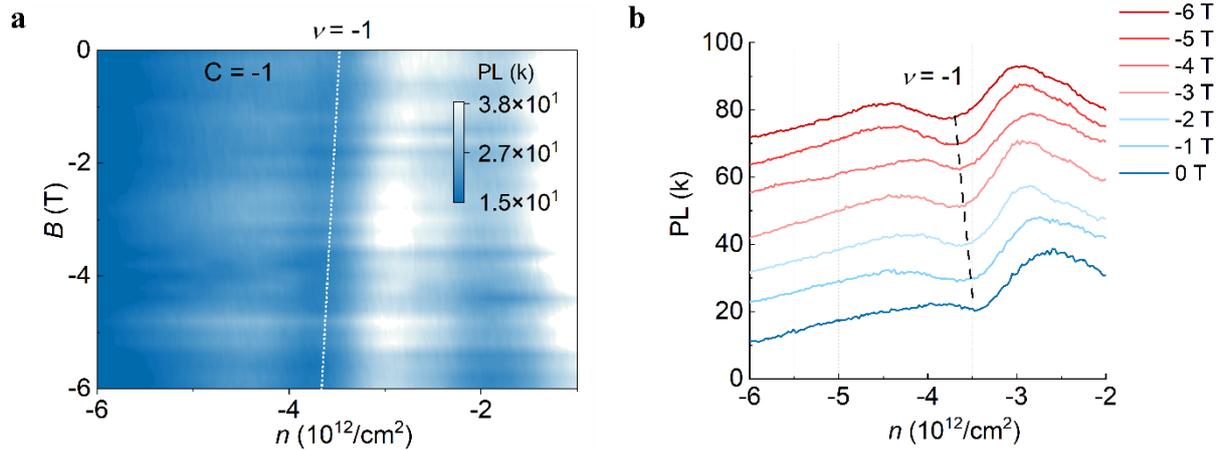

**Supplementary Fig. 5. Integrated PL intensity versus out-of-plane $B$ field and carrier density $n$, Device#O1. a.** The dispersion of the PL counts dip as a function of $B$ and $n$ indicates the formation of Chern insulating state (CI) at filling $v = -1$ in our device. The positions of dips in the incompressible states are extracted from a Gaussian fit as a function of the magnetic field. **b.** The PL versus carrier density $n$ and magnetic field $B$, the magnetic field $B$ is from 0 T to -6 T. The $v = -1$ PL dips are resolved and show linear dispersion with magnetic field $B$.

The topological property is checked by the Streda formula[3]. The filling -1 state shifts linearly with the magnetic field $B$, from 0 T to -6 T. The Chern number is extracted by fitting the linear dispersion of the PL dips in Supplementary Fig. 5a. For filling -1 state we obtain C = -1.3 ± 0.3.

**hBN thickness and carrier density in #Device O1 and #Device O2**

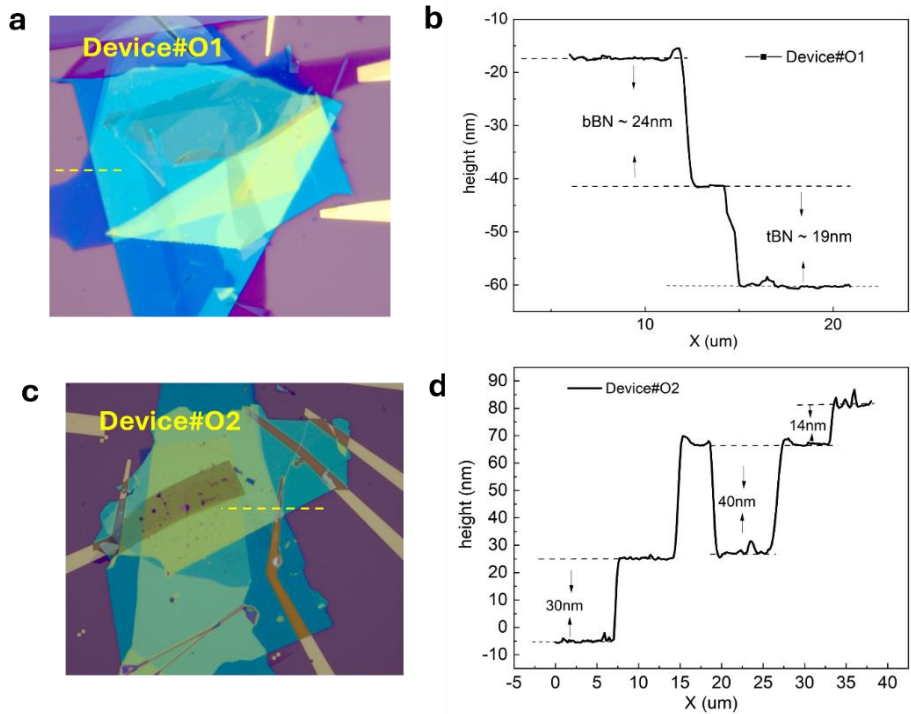

**Supplementary Fig. 6**. **top and bottom hBN thickness of Device a**. In Device#O1, top BN is 19 nm, bottom BN 24 nm. **b**. AFM measurement linecut along the yellow dashed line in a. **c**. In Device#O2, top BN thickness is 44 nm and bottom BN thickness 40 nm. **d**. AFM measurement linecut along the yellow dashed line in c.

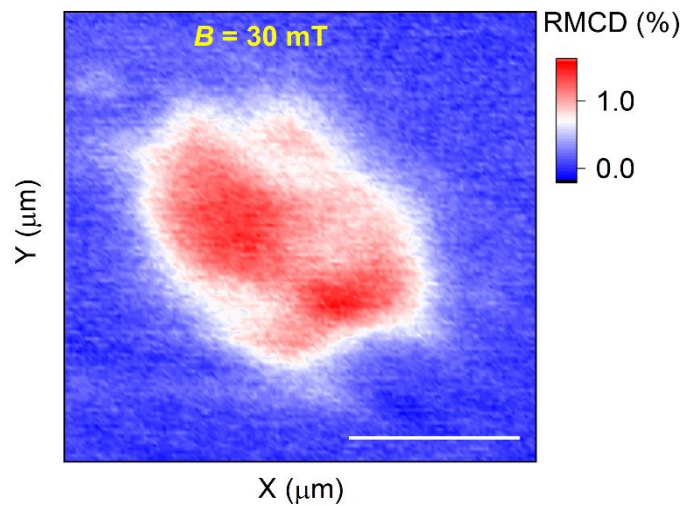

**Supplementary Fig. 7. Ferromagnetic states at filling factor $v = -1.1$ in Device#O1.**

The doping is fixed at filling $v = -1.1$, with $D/\varepsilon_0 \approx 0$ and a small applied out-of-plane magnetic field B = 30 mT. The RMCD signal has a 1 μm real space resolution. The scale bar is 10 μm.

We note that the RMCD signal, with its spatial resolution, can be used to identify areas with homogeneous twist angles[4].

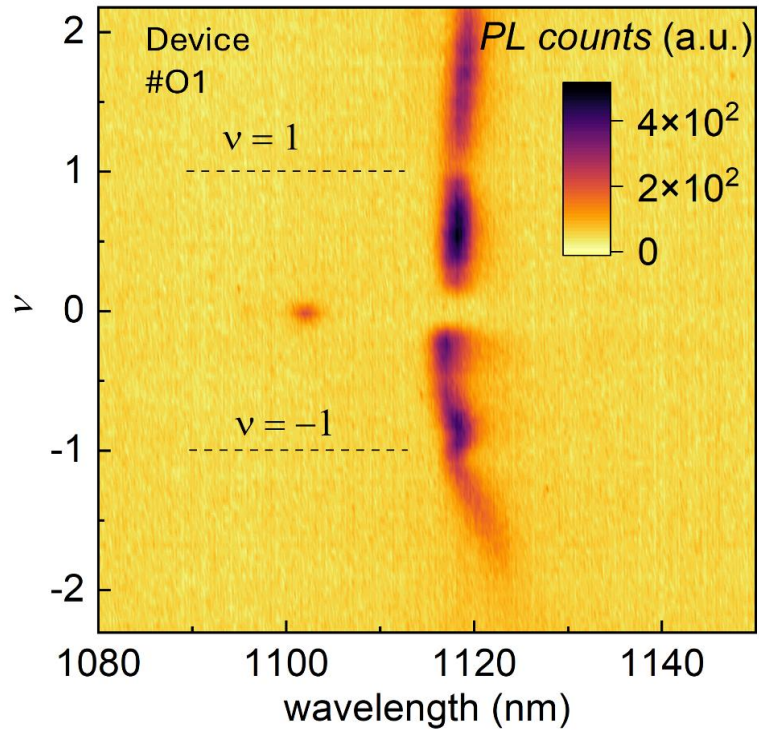

**Supplementary Fig. 8. The PL spectrum of twist bilayer MoTe$_2$.** The PL spectrum of Device #O1, with $D/\varepsilon_0 \approx 0$. A blue shift of the spectrum is detected at the integer filling $v = -1$, similar to previous work[5]. In the electron side, a dip is found at the filling 1 state.

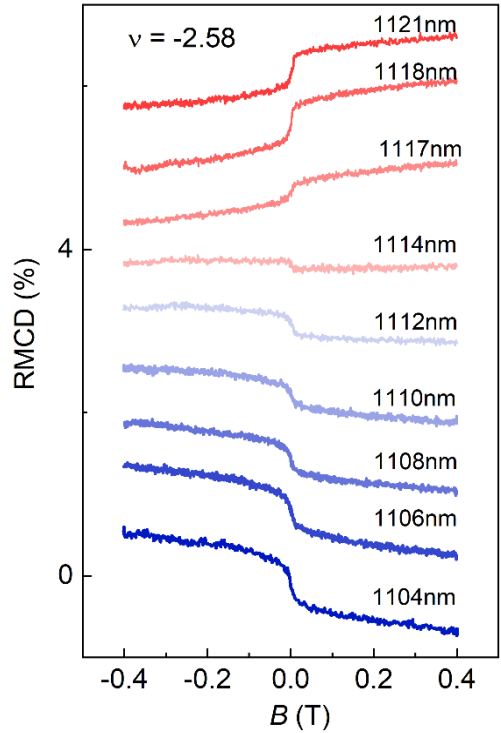

**Supplementary Fig. 9. RMCD wavelength dependence at filling factor -2.58.** The wavelength is set from 1104 nm to 1121 nm, and a sign reversal of the RMCD signal is detected between 1114 nm and 1117 nm.

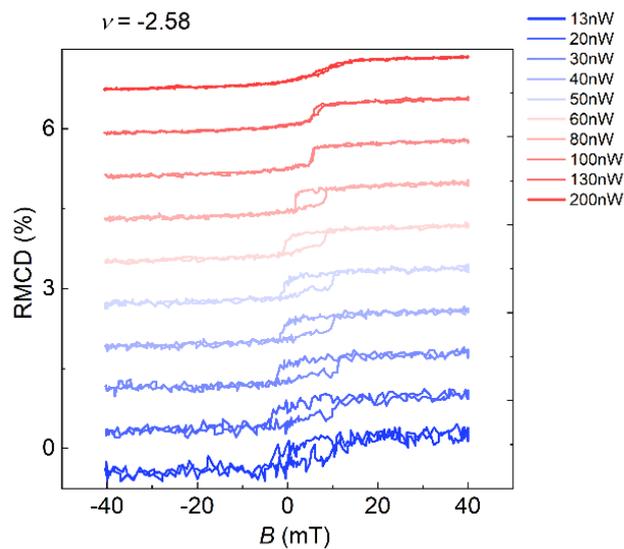

**Supplementary Fig. 10. RMCD laser power dependence at filling factor -2.58.**

We measured the power dependence of the hysteresis loop at the light wavelength of 1120.9 nm. The hysteresis loop gradually disappeared as the power increased from 13 nW to 200 nW. We attribute this effect to the heating from the laser. Given our laser's spot size of approximately 1 μm × 1 μm, high power may have disrupted the magnetic domain due to heating.

## Hysteresis loop of the RMCD versus filling factor near -1

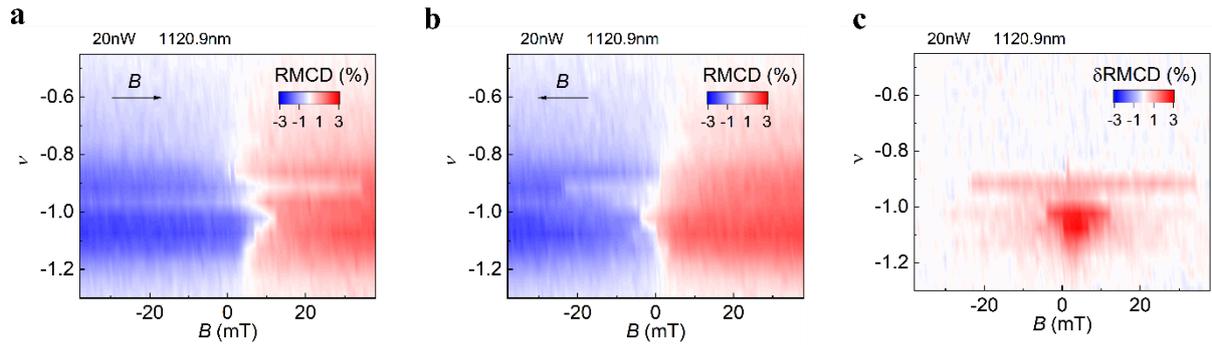

**Supplementary Fig. 11. Doping-dependent ferromagnetic states in the first Moiré band.**
**a.b.** RMCD versus filling and magnetic field swept up and swept down in the first moiré band. **c.** Difference of up swept and down swept, giving the hysteretic component of the RMCD versus filling from -0.45 to -1.3.

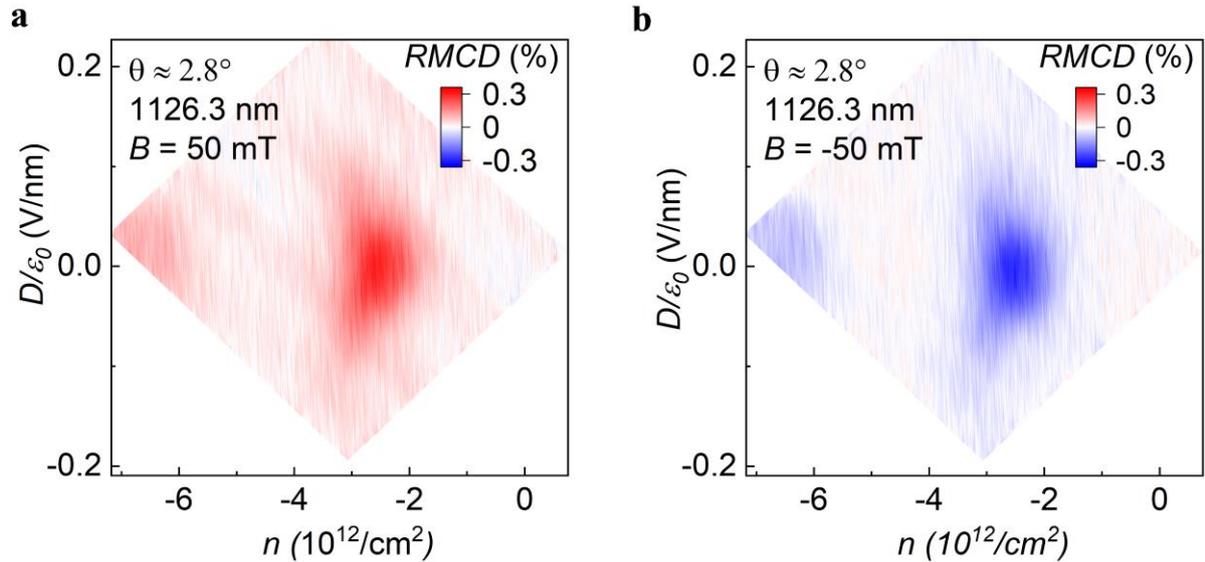

**Supplementary Fig. 12. Ferromagnetism in the first and second Moiré bands in Device#O2. a.** RMCD as a function of the doping filling factor and electric field, measurement wavelength is 1126.3 nm, $B$ is fixed at 50 mT and $T$ = 1.6 K. **b.** RMCD plots as a function of the doping filling factor and electric field, measurement wavelength is 1126.3 nm, B is fixed at -50 mT and T = 1.6 K.

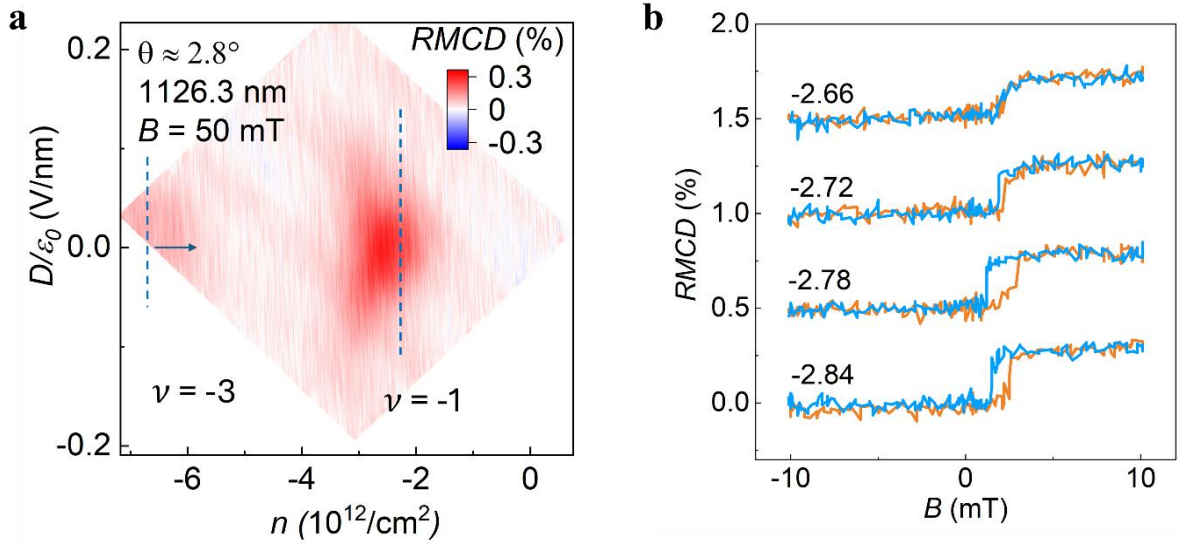

**Supplementary Fig. 13. Doping dependent RMCD in the second Moiré band in Device#O2. a**. RMCD plots as a function of the doping filling factor and electric field. The fillings -1 and -3 are indicated by the dashed line. **b**. Hysteresis loop of the RMCD versus filling factor in the second moiré band, with displacement nearly zero. Measurement is along the arrow line in a**.**

Interestingly, as shown in Fig. 3b and here, an unsymmetric coercive field is observed.

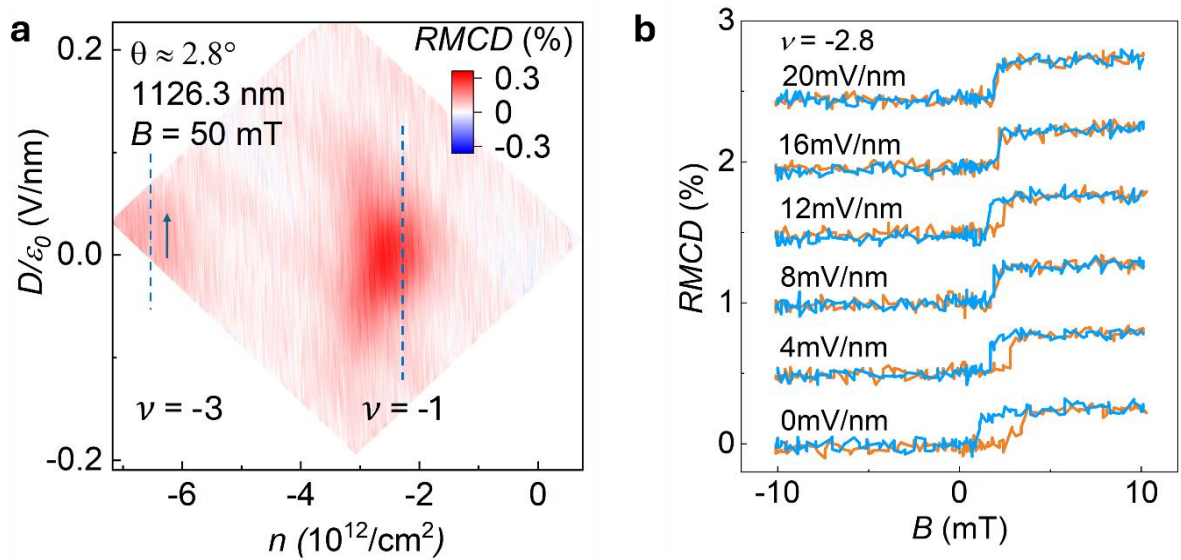

**Supplementary Fig. 14. Electric filed dependent RMCD near filling -2.8 in Device#O2. a.** RMCD map. **b.** Hysteresis loop of the RMCD, versus displacement field at filling $v = -2.8$. Measurement is along the arrow line in a**.**

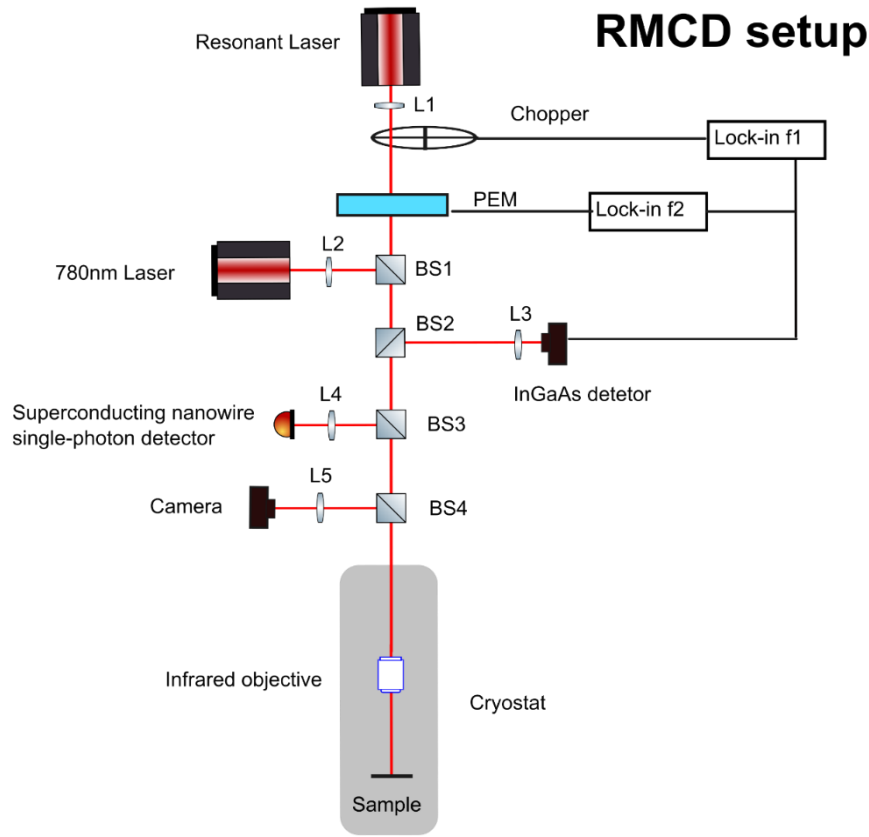

**Supplementary Fig. 15. RMCD set up for the measurement.** The chopper frequency is 912 Hz, PEM frequency is set at 50.5 kHz.